\documentclass[%
 reprint,
superscriptaddress,
 amsmath,amssymb,
 aps,
]{revtex4-2}
\usepackage{soul}
\usepackage{graphicx}% Include figure files
\usepackage{dcolumn}% Align table columns on decimal point
\usepackage{bm}% bold math
\usepackage{amsmath}
\sethlcolor{lightgray}
\usepackage{textalpha}
\usepackage{comment}
\usepackage{color}

\begin{document}

\preprint{APS/123-QED}
\title{The hidden hierarchical nature of soft particulate gels}

\author{Minaspi Bantawa}
\affiliation{Department  of  Physics,  Institute  for Soft  Matter  Synthesis and Metrology,
Georgetown  University,  37th and O Streets,  N.W., Washington,  D.C. 20057,  USA}

\author{Bavand Keshavarz}
\affiliation{Department of Mechanical Engineering, Massachusetts Institute of Technology,
77 Massachusetts Avenue, Cambridge, Massachusetts 02139, USA
}

\author{Michela Geri}
\affiliation{Department of Mechanical Engineering, Massachusetts Institute of Technology,
77 Massachusetts Avenue, Cambridge, Massachusetts 02139, USA
}
\author{Mehdi Bouzid}
\affiliation{Univ. Grenoble Alpes, CNRS, Grenoble INP, 3SR, F-38000 Grenoble, France}

\author{Thibaut Divoux}
\affiliation{ENSL, CNRS, Laboratoire de physique, F-69342 Lyon, France}

\author{Gareth H. McKinley} 
\affiliation{Department of Mechanical Engineering, Massachusetts Institute of Technology,
77 Massachusetts Avenue, Cambridge, Massachusetts 02139, USA
}
\author{Emanuela Del Gado}
\affiliation{Department  of  Physics,  Institute  for Soft  Matter  Synthesis and Metrology,
Georgetown  University,  37th and O Streets,  N.W., Washington,  D.C. 20057,  USA}

\date{\today}
\begin{abstract}
Soft particulate gels include materials we can eat, squeeze, or 3D print. From foods to bio-inks to cement hydrates, these gels are composed of a small amount of particulate matter (proteins, polymers, colloidal particles, or agglomerates of various origins) embedded in a continuous fluid phase. The solid components assemble to form a porous matrix, providing rigidity and control of the mechanical response, despite being the minority constituent. The rheological response and gel elasticity are direct functions of the particle volume fraction $\phi$: however, the diverse range of different functional dependencies reported experimentally has, to date, challenged efforts to identify general scaling laws. Here we reveal a hidden hierarchical organization of fractal elements that controls the viscoelastic spectrum, and which is associated with the spatial heterogeneity of the solid matrix topology. The fractal elements form the foundations of a viscoelastic master curve, which we construct using large-scale 3D microscopic simulations of model gels, and can be described by a recursive rheological ladder model  over a range of particle volume fractions and gelation rates. The hierarchy of the fractal elements provides the missing general framework required to predict the gel elasticity and the viscoelastic response of these ubiquitous complex materials. 
\end{abstract}

\maketitle

\section{Introduction}

 For gels formed through polymerization reactions or crosslinking of polymers in solution, 80 years of polymer physics have provided the basis to fully understand the links between chemical architectures and rheology \cite{Flory1953,deGennes-book}. Percolation theory has been central for understanding the gel properties as a function of the distance from a gelation (percolation) threshold \cite{stauffer1982gelation}. The self-similarity of the chemical architectures close to the percolation threshold naturally produces a hierarchy of lengthscales and timescales, leading to power-law characteristics in the viscoelastic response \cite{daoud2000,WinterChambon1986,Muthukumar1989,Martin1988a,Adolf1990}. By contrast, in particulate gels, the link between microstructure and viscoelasticity remains elusive. Such gels can be formed from both synthetic or natural constituents, and represent a preferred strategy to incorporate high-value functional components while limiting costs and risks. These gels form through physical association of the initial colloidal building blocks, due to surface forces and attractive interactions mediated by the solvent \cite{Witten-book,TrappePhase2001, Lu2008,petekidis_wagner_2021,royall2021real}. Ultimately, they develop as non-equilibrium structures produced by frustration in the growth of aggregates, interconnected and locked into larger-scale disordered assemblies, from which rigidity and viscoelasticity emerge. There is growing evidence that in this class of gels a percolation threshold may also universally control the onset of rigidity (rigidity percolation) and gel elasticity \cite{Zhang2019,Whitaker2019,Tsurusawa2019}; however, the microscopic origin of that percolation transition and of the resulting power-law rheological response, observed over a range of compositions and solid contents, remain unclear. The extreme variability of gel microstructures \cite{Dinsmore_2002, Lu2008, Whitaker2019, royall2021real} and microscopic dynamics \cite{Duri:2006, Szakasits2017dynamics, Negi2014, Aime2018, KeshavarzPNAS2021} revealed by experiments seems to suggest that the microscopic physical origins of the macroscopic rheological response need to be established on a case-by-case basis. The particle volume fraction $\phi$ is the main control parameter in experiments, which invariably report a strongly varying shear modulus $G_{0} \propto \phi^{f_{\textrm{obs}}}$, however $f_{\textrm{obs}}$ ranges widely from $3$ to $8$, again questioning the existence of any universal behavior and of a general framework to predict the mechanical response \cite{Shih1990, TrappePhase2001, Buscall1988, Russel1993, Piau1999, Trappe2000, Prasad2003, Wyss2005, Yanez1996, Ramakrishnan2004, Mellema2002}.

\begin{figure*}[htb]
\centering
\includegraphics[scale=0.97]{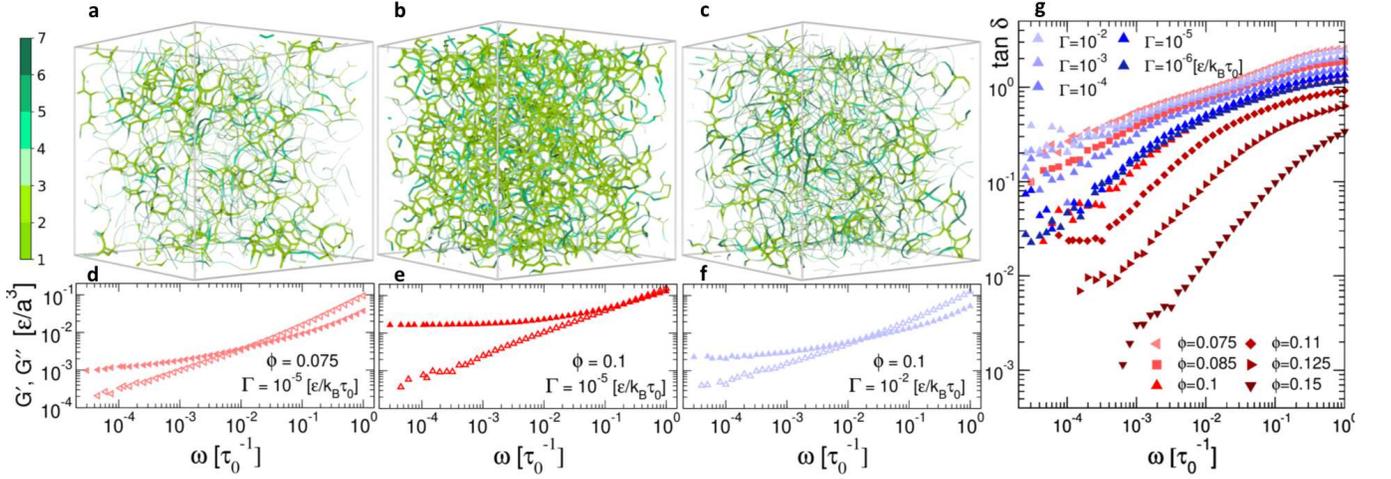}
\caption{\textbf{Microstructural and viscoelastic properties of networks having different connectivity.} Simulation snapshots of gel networks at different volume fractions (a) $\phi=0.075$ and (b) $\phi=0.1$, prepared with a fixed gelation rate $\Gamma=10^{-5} \epsilon /k_B \tau_0 $, and (c), $\phi=0.1$ with $\Gamma=10^{-2} \epsilon /k_B \tau_0$. The microstructural snapshots are colored based on the mesh size, i.e., the topological distance between branching points and the thickness of the bonds is proportional to the local density of branching points. The corresponding viscoelastic spectra are shown in (d)-(f) where the filled and open symbols represent the storage modulus $G^\prime$ and loss modulus $G^{\prime\prime}$, respectively. (g) Variation in the loss tangent ($\tan \delta = G''/G'$) vs.~frequency for gels with different volume fractions $\phi$ and gelation rates $\Gamma$.}\label{fig1}
\end{figure*}

\section{Results and Discussion}
\textbf{Gel microstructures, viscoelasticity and rheological master curve.} 
We use 3D numerical simulations of a particle-based model that capture the microscopic dynamics and rheology of soft particulate gels \cite{Colombo:2014SM,Colombo:2014JOR,Bouzid2017,Bouzid2018,Bantawa2021} (see also Methods). In terms of general trends, for a given gelation rate $\Gamma$, increasing the solid volume fraction $\phi$ increases, on average, local connectivity and gel elasticity, by increasing the amount of branching in the gel [Figs.~\ref{fig1}(a),(b)]. For a given $\phi$, reducing the gelation rate also favors the  branching of strands as the network self-assembles, leading to structures with higher local connectivity and elasticity [Figs.~\ref{fig1}(b),(c)]. However, gels formed at lower $\phi$ are more sparsely connected and their local connectivity is also more spatially heterogeneous [Fig.~\ref{fig1}(a)].

For all gels, the linear viscoelastic spectra $G'(\omega)$ and $G''(\omega)$ [Fig.~\ref{fig1}(d)-(f)] are computed using the OWCh protocol \cite{Geri2018,Bouzid2018}, which yields fast and accurate estimates of 
the mechanical properties over a wide range of deformation frequencies, and we use reduced simulation units to scale both moduli and frequency (see Methods). As in experiments \cite{Trappe2000,Prasad2003,Negi2014,Helgeson2014,Rao2019,Huang2021}, varying the particle volume fractions $\phi$ over a relatively small range (i.e., between $5\%$ and $15\%$) produces apparently minor changes in the microstructure but translates into dramatic variations of the viscoelastic strength and characteristic timescales [Fig.~\ref{fig1}(d),(e)]. Changing the gelation rate for a fixed $\phi$ leads to similar observations [see Fig.~\ref{fig1}(e),(f) and Fig.~S1].    

The frequency dependence of the loss tangent $\tan \delta = G''/G'$ [Fig.~\ref{fig1}(g)] summarizes the mechanical response of 11 gels, obtained for different $\phi$ and $\Gamma$. In spite of the wide range of driving frequencies, all of the data sets are broadly self-similar and slowly approach a high-frequency plateau.
\begin{figure}[htb]
\centering
\includegraphics[scale=1.0]{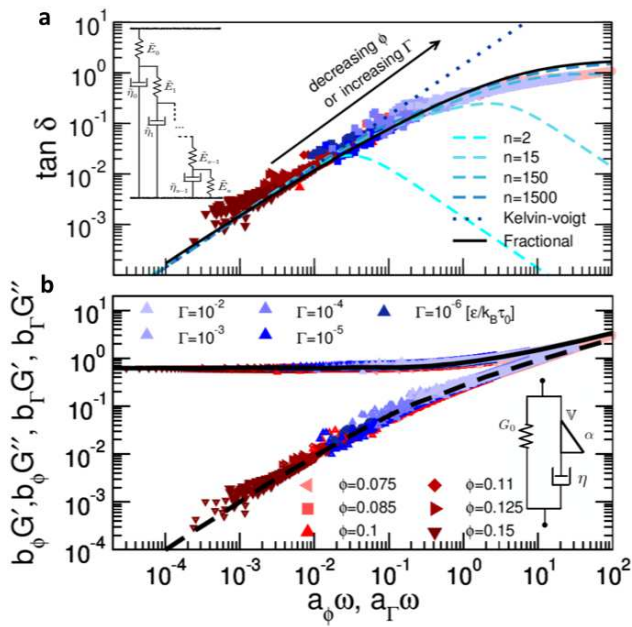}
\caption{\textbf{Rheological master curves and constitutive model for particulate gels.} (a) Superposition of the self-similar curves of $\tan \delta$ onto a single master curve achieved by rescaling the imposed deformation frequency with a shift factor $a_\phi$ or $a_\Gamma$ w.r.t.~a reference gel at a volume fraction $\phi=0.15$. The functional form of $\tan \delta$ predicted by ladder models with increasing number of elements $n$, as well as by the asymptotic fractional model are shown by different lines as listed in the legend. The inset shows a schematic representation of the corresponding ladder model. (b) 
Master curves for the moduli with the same horizontal shift factors $a_\phi$ or $a_\Gamma$ and vertical shift factors $b_\phi$ or $b_\Gamma$.  
The solid and dashed lines represent, respectively, the predictions for $G^{\prime \prime}$ and $G^\prime$ with the four parameter fractional model shown in the inset.
}\label{fig2}
\end{figure}
An horizontal shift, rescaling the frequency either by a factor $a_{\phi}$ at a given $\Gamma$, or by a factor $a_{\Gamma}$ at fixed $\phi$, leads to a unique master curve for $\tan \delta$, covering six decades of rescaled frequency [Fig.~\ref{fig2}(a)]. Here we use $\phi = 15\%$ and $\Gamma=10^{-5} \epsilon /k_B \tau_0 $ as the reference conditions for collapsing the data.

\textbf{Ladder and fractional models.} 
The resulting master curve exhibits an extended power-law regime, highlighting a hierarchy of timescales that 
is captured by recursively combining viscoelastic elements 
in a hierarchical ladder structure \cite{Schiessel1993,Schiessel1995c}. The ladder-like arrangement, sketched as an inset in Fig.~\ref{fig2}(a), comprises $n$ viscoelastic elements with model contributions $(\tilde{E_i},\tilde{\eta_i})$ (with $0\leq i \leq n$) and an exponent $\alpha$ that sets the relationship between $(\tilde{E_i},\tilde{\eta_i})$ and $(\tilde{E_0},\tilde{\eta_0})$ [see Eq.~\eqref{recur} in Methods]. For large $n$ ($n \ge 150$), the ladder model predicts a loss factor $\tan\delta$ that smoothly transitions from a linear increase to a plateau at high frequencies (see also Methods), in good agreement with the master curve obtained from the simulation data [Fig.~\ref{fig2}(a)]. We can now vertically rescale loss and storage moduli by a factor $b_\phi$ (or $b_\Gamma$) to obtain master curves for $G'$ and $G''$ as shown in Fig.~\ref{fig2}(b). Taking the continuous limit of the ladder model introduced in Fig.~\ref{fig2}(a), we obtain a more compact description of the viscoelastic response in terms of a \textit{fractional} Kelvin-Voigt model characterized by just four parameters: a spring constant ($G_0$), a viscous dashpot ($\eta$), and a fractional element or `spring-pot' (characterized by a scale factor $\mathbb{V}$ and an exponent $\alpha$) \cite{Jaishankar:2013}. The power-law exponent $0\leq\alpha\leq1$ reflects the recursive nature of the underlying ladder model, and we can relate the other parameters to the rungs of the ladder model in the limit $n\rightarrow \infty$ (see SI Sections 2 \& 3):
\begin{equation}\label{laddertofrac}
   G_0=\frac{\tilde{E_0}}{{n}^{2\alpha}},\:\eta={n}^{2-2\alpha}\tilde{\eta_0},\:\mathbb{V}=\tilde{E_0}\left(\frac{\tilde{\eta_0}}{\tilde{E_0}}\right)^\alpha.
\end{equation}

 The extended power-law regime evident in the master curves and its description by a ladder model reflect the scale-free characteristics of the relaxation spectra underpinning the viscoelastic response For polymer gels, power-law characteristics and rheological ladder models directly stem from the self-similar chemical architecture close to percolation 
 \cite{DeGennes1974b, Stauffer1976,Stauffer2018,daoud2000}. In soft particulate gels, instead, the microstructures are often not self-similar \cite{TrappePhase2001,Dinsmore_2002,Lu2008,Whitaker2019,Negi2014}, as is also the case here (see Fig.~S3). Moreover, {\it both} $\phi$ and $\Gamma$ determine the range of frequencies and viscoelasticity relevant to the power-law region of the spectra [Fig.~S1(c)], demonstrating the intricate coupling between particle volume fraction and gelation kinetics, which makes the microstructural origin of the rheology of this class of gels so difficult to pin down.

\begin{figure*}[htb]
\centering
\includegraphics[width=1.0\textwidth]{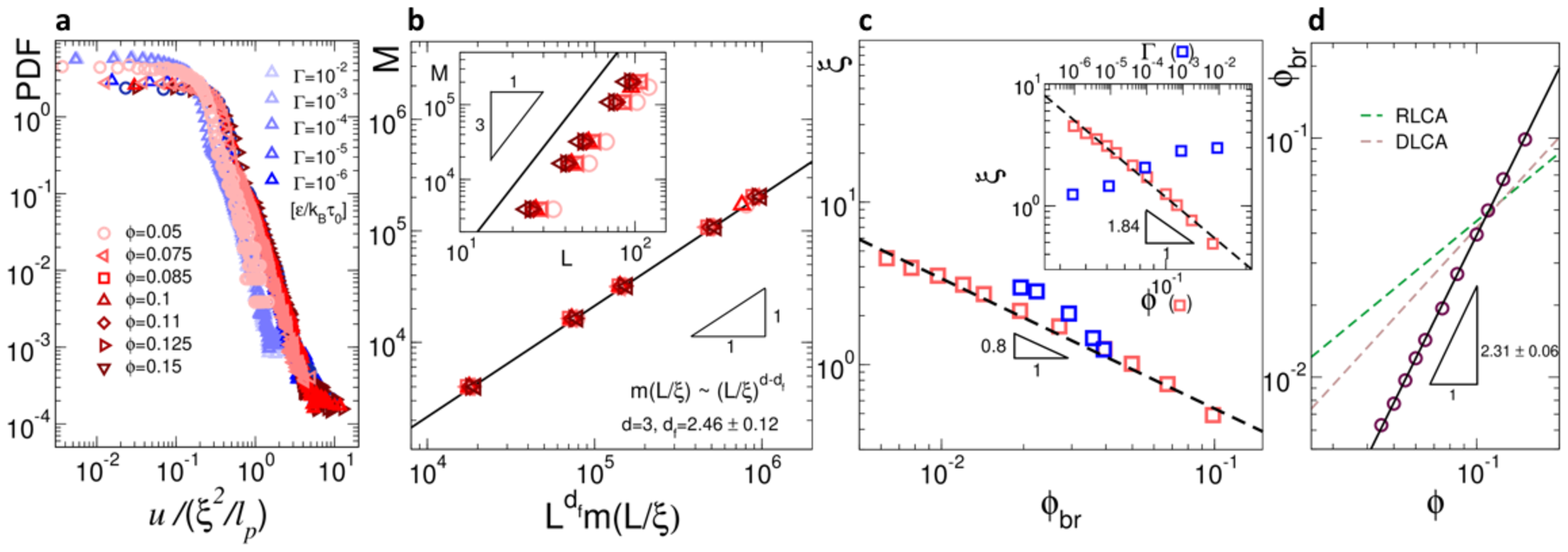}
\caption{\textbf{Structural characterization of gel networks.} 
(a) Distributions of fluctuations of displacements $u=[({\bf \Delta} - \langle {\bf \Delta} \rangle)^2]^{1/2}$ normalized by $\xi^2/l_p$. (b) Master curve for the mass $M$ of the gel network vs normalized system size $L^{d_f}m(L/\xi)$ for different volume fractions, where $m(L/\xi)=(L/\xi)^{d-d_f}$ with a fractal dimension $d_f=2.46\pm0.12$. The data is obtained by changing the system size for different volume fractions. Inset: $M$ vs the system size $L$. (c) Correlation length $\xi=\langle (l - \langle l \rangle)^2\rangle^{1/2}$. The red (resp.~blue) data correspond to various volume fractions $\phi$ (resp.~gelation rate $\Gamma$). The dashed line is a power law of exponent $-0.8$. Inset: Evolution in correlation length $\xi$ vs.~volume fraction $\phi$ (bottom axis and red symbols) and vs.~gelation rate (top axis and blue symbols).
 (d) Volume fraction of branching points $\phi_\mathrm{br}$ vs.~the volume fraction of particles $\phi$: circles correspond to the simulations data. The continuous line shows the best fit of the data by a power law of exponent $2.31\pm0.06$. Corresponding predictions for the DLCA and RLCA scenarios are shown by dashed lines of slope $1.00\pm0.06$ and $1.30\pm0.06$, respectively. 
}\label{fig3}
\end{figure*}

\begin{figure*}[htb]
\centering
\includegraphics[scale=0.95]{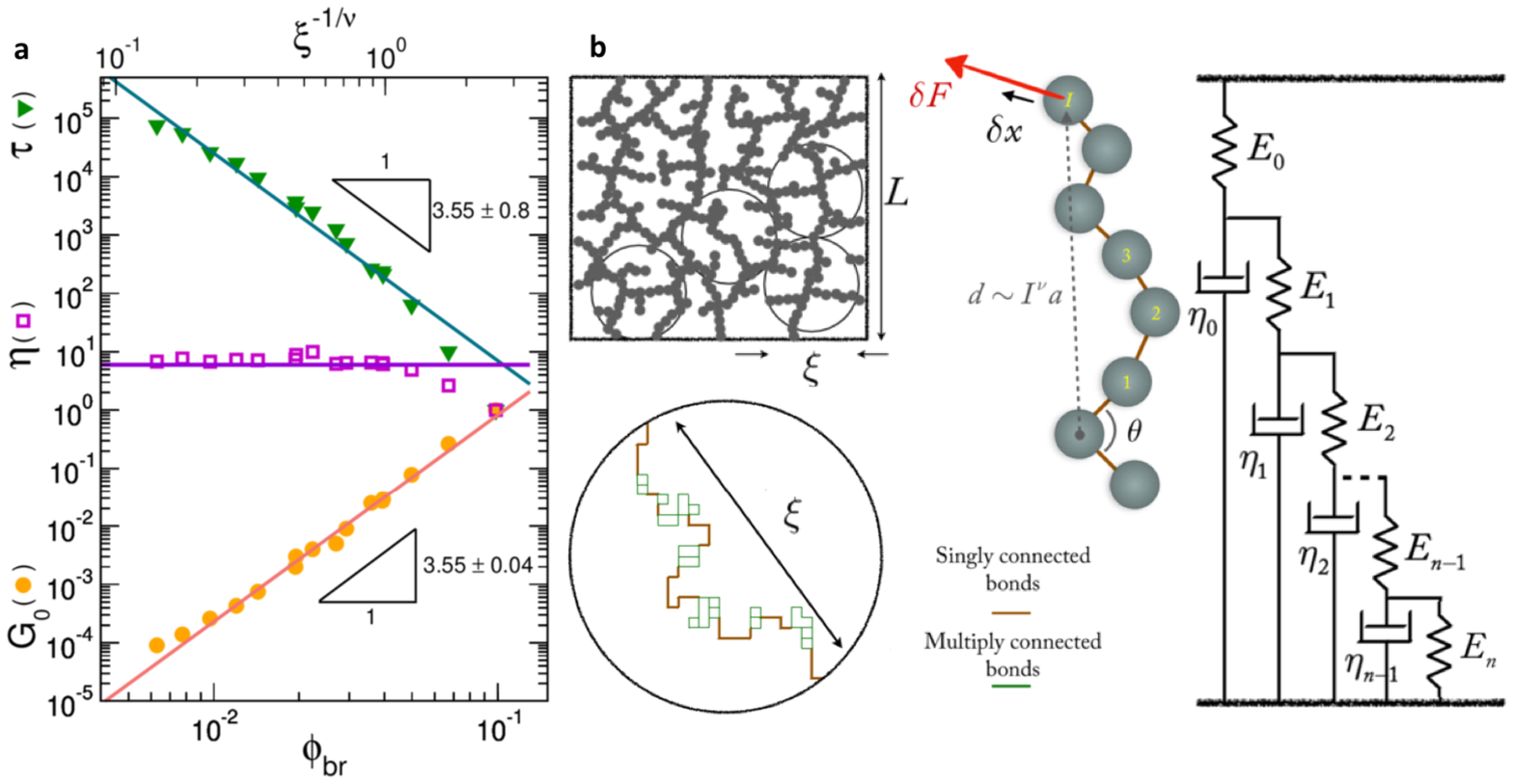}
\caption{ \textbf{Scaling of viscoelastic parameters and connection to the ladder model.} (a) Scaling of retardation time $\tau\sim$ $a_\phi$ or $a_\Gamma$ (triangles), elastic modulus $G_0\sim 1/b_\phi$ or $1/b_\Gamma$  (circles) and  characteristic viscosity $\eta = \tau G_0$ (squares) vs  volume fraction of branching points $\phi_\textrm{br}$ (bottom axis) and $\xi^{-1/\nu}$ (top axis). (b) Schematic representations of the underlying ladder model: fractal blobs of size given by the correlation length $\xi$ (top left), singly and multiply connected bonds within a single blob (bottom left), bending moment of gel strands for which bond bending costs energy when deformation is applied (middle), and the equivalent heirarchical mechanical element representation, consisting of multiple springs and dashpots (right).}
\label{fig4}
\end{figure*}

\textbf{Lengthscale and fractal characteristics.} To tease out the microscopic origin of the rheological response, 
we analyze the fluctuations in the spontaneous microscopic dynamics across all our gels at rest and subjected only to thermal fluctuations (see Methods). In the simulations we can use a suitable range of thermal fluctuations such that no significant changes in the gel structures are induced over the simulation time window. We then compute the displacements ${\bf\Delta}={\bf{r}}(t_0+t_w)-{\bf{r}}(t_0)$ from the particle trajectories ${\bf{r}}(t)\equiv(x(t),y(t),z(t))$, where the time interval $[t_{0}$, $t_{0}+t_{w}]$ is such that both $t_{0}$ and $t_{w}$ are in the plateau region of the particle mean-squared displacement as a function of time [see Methods and Fig.~S4(a) in SI]. 
 The fluctuations $u=[({\bf \Delta} - \langle {\bf \Delta} \rangle)^2]^{1/2}$ are widely distributed across the different gel microstructures [see Fig.~S4(b)]. 
In complex gel architectures, both microscopic dynamics and mechanics are largely controlled by the presence of more densely connected regions interspersed with sparsely connected ones \cite{Colombo:2013,Bouzid2017,Bouzid:2018LAN, shivers2019scaling,Burla2019from}. We therefore measure, along network strands, the distribution of {\it topological} distances $l$ which separate two connected branching points. This distribution provides direct access to the structural and micromechanical heterogeneities in the gels that determine floppy modes and low frequency elasticity \cite{Rocklin2021elasticity}. From the probability distribution $p(l)$ across all gels (Fig.~S5), we extract the variance and use $\xi=\langle ( l - \langle l \rangle)^2\rangle^{1/2}$, with dimensions of length, to characterize the gel mesh size heterogeneity. By rescaling all fluctuations $u$ of the microscopic displacements with $\xi^2/l_{\rm p}$, where $l_{\rm p}$ denotes the persistence length of the gel strands \cite{Bantawa2021}, the distributions $p(u)$ collected across all gels collapse onto a unique curve [Fig.~\ref{fig3}(a)]. Hence the variation of $\xi$ with $\phi$ and $\Gamma$ captures the microstructural origin of the variations in the microscopic dynamics.

 As the gels become softer with decreasing $\phi$ or increasing $\Gamma$, less connected networks are produced, and $\xi$ grows because less connected networks are also more spatially heterogeneous. Extrapolating, at the very onset of rigidity, $\xi$ captures the first rigid backbone, a single branch that spans the whole gel and that is sufficient, alone, to provide rigidity. These considerations point to $\xi$ as a direct probe of the distance from the rigidity threshold in our gels. Such a metric, in fact, is ultimately set by the number of branching points, which we can measure, in our model gels, through the volume fraction $\phi_\textrm{br}$ of particles with coordination $z$ = $3$. While $\xi$ varies with both $\phi$ and $\Gamma$, the data across all gels follow the scaling $\xi \propto \phi_\textrm{br}^{-\nu}$ [Fig.~\ref{fig3}(c)], suggesting that indeed $\xi$ may capture the scaling of the critical correlation length associated with the rigidity transition that governs the emerging gel elasticity. The computed estimate for $\nu \simeq 0.80 \pm 0.16$ is compatible with the value of a 3D random percolation network \cite{Stauffer2018}. Since we verify from the plateau in $G^\prime$ at low frequencies that all our gels are rigid, the power-law dependence of $\xi$ on $\phi_\textrm{br}$ may reflect that, because of their extreme softness and structural complexity, soft particulate gels are marginally rigid and remain relatively close to a rigidity percolation threshold over a range of particle volume fractions. 

 If the rigidity percolation transition in particulate gels is akin to a random percolation, then following the blob-links-nodes model for the self-similar structure of a spanning cluster in percolation theory \cite{DeGennes1974b,Stauffer2018,Coniglio1981}, each particulate gel is, effectively, a disordered network composed of fractal elements (blobs) whose linear size scales with $\xi$ and whose fractal dimension is $d_{f}$. For a gel sample of linear size $L$, the volume $L^d$ (in $d$ dimensions) will consist of $\left(L/\xi\right)^d$ sub-boxes, each containing a mass fraction of the gel $\propto \xi^{d_f}$. Close to the percolation threshold, the scaling hypothesis for a critical point dictates that 
 $M(L,\xi) \propto \xi^{d_f}m\left(L/\xi\right)$, where the scaling function is $m\left(L/\xi\right)= \left(L/\xi\right)^{d-d_f}$ \cite{Stauffer2018}.
 
When we compute the gel mass for samples with a range of sizes $L$ for each volume fraction (here $L$ is the linear size of the simulation box), the data are spread out and grow as $L^{3}$ [inset in Fig.~\ref{fig3}(b)]. However, if we use the scaling argument just laid out, all data collapse onto the unique scaling function $m\left(L/\xi\right)$ developed above [Fig.~\ref{fig3}(b)] for $d_{f}\simeq 2.46 \pm 0.12$, a fractal dimension again consistent with a 3D random percolation network (see Fig.~S7(b)). Inferring the frequency dependence of the viscoelastic modulus just from $d_{f}$ as $\alpha=d/(d_f+2)$, as proposed for polymer networks \cite{Muthukumar1989,KeshavarzPNAS2021,Aime2018}, yields $\alpha\simeq0.67$ in good agreement with the predictions of the fractional and ladder models ($\alpha=0.66 \pm 0.05$) for the viscoelastic master curves [Fig.~\ref{fig2}(a),(b)]. 

Following further the blob-links-nodes model, each fractal element should contain loops and {\it singly connected bonds}, whose number $N_{\rm SCB}$ diverges, as $\xi$ also does, at the percolation threshold ($N_{\rm SCB}\propto \xi^{1/\nu}$). Indeed, close enough to the threshold, singly connected bonds should be present at all lengthscales, and organized in a self-similar fashion \cite{Coniglio1981}. This implies that $N_{\rm SCB} \propto 1/\phi_\textrm{br}$, and that $\phi_\textrm{br}$ contains the information on how singly connected bonds become progressively more prevalent, over all lengthscales, as $\xi \longrightarrow L$ (and $\phi_\textrm{br} \longrightarrow 0$). Hence the fact that $\xi$ and $\phi_\textrm{br}$ control both the microscopic dynamics and the bulk rheology of our gels can be directly related to the hierarchical organization of the singly connected structures. 

The fractal blobs whose linear size $\propto \xi$ and with fractal dimension $d_{f}$ fill the gel volume for any $\phi_\textrm{br} \neq 0$, hence $\xi \propto \phi^{-1/(d-d_{f})}$ and, combining with $\xi \propto \phi_\textrm{br}^{-\nu}$, we obtain  $\phi_\textrm{br} \propto \phi^{1/\nu(d-d_{f})}$. The simulation results satisfy this scaling prediction, if we use $\nu \simeq 0.8$ and $d_{f} \simeq 2.5$ as obtained previously from our data [Fig.~\ref{fig3}(d)].  
We note that, if these fractal elements controlling the rheology were the fractal aggregates formed through diffusion-limited or reaction-limited cluster aggregation (respectively DLCA or RLCA) quite common in colloidal suspensions, their fractal dimensions would be different (respectively $d_{f} \approx 1.8$ or $\approx 2.1$ \cite{LinFractalDimension1989,Shih1990,Witten-book}) and this would lead to markedly different scalings between $\phi_\textrm{br}$ and $\phi$ [cf. Fig.~\ref{fig3}(d)]. For DLCA aggregates our scaling %approximately 
translates into $\phi \propto \phi_\textrm{br}$, with the particle volume fraction directly setting the distance from the rigidity threshold, consistent with the analysis of fractal aggregation in colloidal gels \cite{Weitz1985, Witten-book}. The aggregation process considered here corresponds to a more general case, as density fluctuations and collective microscopic dynamics contribute to the microstructure development \cite{Zhang2019}, and may apply, at a coarse grained level, to a broader range of particulate gels \cite{Lu2008,Whitaker2019,Rocklin2021elasticity,Wyss2005}.

\textbf{Elastic percolating network.} 
We now consider the mechanics of fractal elements of linear size $\propto \xi$ having an elastic stiffness $K_\xi$. Assuming that they are uniformly distributed in space, the resulting elastic stiffness of the gel can be estimated as $K \propto (L/\xi)^{d-2}K_\xi$. With bending elasticity \cite{Kantor1984}, $K_\xi$ directly depends on the presence of singly connected bonds $K_\xi=K_0/(N_{\rm SCB}\xi^2)$, where $K_0$ is the torsional bending stiffness between neighboring bonds, which, in our case, can be computed from the microscopic interactions \cite{Bantawa2021}. 
Identifying the rigidity transition with random percolation as demonstrated above, close enough to the threshold, the gel modulus $G_{0}$ should scale with $\xi$  as 
\begin{equation}\label{KW eqn}
   G_0\propto \xi^{-f/\nu} 
\end{equation} where $f=\nu d +1$ \cite{Kantor1984}. Based on Fig.~\ref{fig3} and the related discussion, these theoretical scaling predictions imply that $G_{0} \propto \phi_{\rm br}^{f}$. For $\nu\simeq 0.8$ we find that $f \simeq 3.5$ in 3D. The scaling that we measure as a function of $\phi_\textrm{br}$ (or $\xi$) from the low-frequency shear modulus $G_0$ of our gels, which also coincides with the vertical shift factor $b$ in our master curves [see Fig.~S8], matches well with this prediction ($f=3.55\pm 0.04$) [Fig.~\ref{fig4}(a)]. We note that in the %specific 
case of DLCA aggregates constituting the fractal elements responsible for rigidity, since $\phi \propto \phi_\textrm{br}$ [Fig.~\ref{fig3}(d)], we obtain $G_{0} \propto \phi^{f}$ and $f \simeq 3.5$, in agreement with the behavior typically found in colloidal gels where diffusion-limited aggregation processes form the initial fractal flocs \cite{Weitz1985,Shih1990}. Our analysis therefore highlights how the dependence of $\phi_\textrm{br}$ (which measures the distance from the rigidity threshold) on $\phi$ (the actual particle content) changes with the specific aggregation process at play [see three examples in Fig.~\ref{fig3}(d)]. In experiments, however, typically only $\phi$ is directly controllable, from which a general dependence $G_{0} \propto \phi^{f_{\textrm{obs}}}$ can be extracted. Hence, while the rigidity percolation transition remains universal to particulate gels, we obtain 

\begin{equation}
f_{\textrm{obs}} = f/\nu (d-d_{f}),
\end{equation}
which naturally has a range of values depending on the fractal dimension of the gel $d_{f}$ (as reflected in Table \ref{table1}), shedding light onto a wide range of experimental observations \cite{TrappePhase2001,Shih1990,Buscall1988,Grant1993,Piau1999,Trappe2000,Prasad2003,Wyss2005,Yanez1996,Ramakrishnan2004,Mellema2002} (see also SI). 

\begin{table}[h!]
\centering
\renewcommand*{\arraystretch}{1.2}
\caption{Scaling exponents for the evolution of the gel modulus  with $G_0 \propto \phi_{\rm br}^f$ and $G_0 \propto \phi^{f_{\rm obs}}$ for different colloidal aggregation processes. The observable power-law exponent $f_{\rm obs}$ is related to the power law $f$ by the expression $f_{\textrm{obs}} = f/\nu (d-d_{f})$, where $d=3$, $\nu=0.8$ and $d_f$ is the fractal dimension.}
\label{table1}
\begin{tabular}{|c|c|c|c|}
\hline
Aggregation process & $d_f$  & $f$ & $f_{\textrm{obs}}$\\
\hline
DLCA & $1.8$ & $3.5$ & $3.6$\\
RLCA & $2.1$  & $3.5$ & $4.8$\\
Present work & $2.5$  & $3.5$ & $8.1$\\
\hline
\end{tabular}
\end{table}

Finally, the horizontal shift factors in our master curves [Figs.~\ref{fig2}(b) and \ref{fig2}(c)] identify a characteristic timescale $\tau$, which we can trace back to the delay time for the gel elastic response to emerge from the microscopic fluctuations [see Fig.~S8] and follows the same scaling as the elastic modulus $G_{0}$. This result, which can be tested in microrheology experiments, explains why the viscous element of the fractional model remains essentially constant for all of the gels [Fig.~\ref{fig4}(a)]. 
The scaling of $\tau$ with the critical lengthscale $\propto \xi$ that describes the fractal blobs (and $\phi_\textrm{br}$) is yet another strong signature of how the topological-dependence of the gel modulus and heterogeneity determine the relaxation spectra. 
 
\textbf{From fractal characteristics to a hierarchical ladder model.} We now demonstrate that the fractal blobs, uniformly distributed in $d$-dimensions, indeed give rise to a mechanical ladder model [Fig.~\ref{fig4}(b)], as hypothesized in Fig.~\ref{fig2}. This model results in a compact description characterized by the four microscopic parameters ${E_0},\eta_0,n$, and $\alpha$, and the overall mechanical response can be obtained as:
\begin{equation}\label{laddertofrac2}
   \left(\tilde{E_0},\tilde{\eta_0},\tilde{\mathbb{V}}\right)=\left(\frac{L}{\xi}\right)^{d-2}\left(E_0,\eta_0,\mathbb{V}\right).
\end{equation} 
where $(L/\xi)^{d-2}$ is a purely geometrical factor.
The elasticity of the ladder structures is set by the bending stiffness of the gel strands $E_0\propto K_0/a^3$ where $K_0$ is the torsional stiffness, with dimension of [force $\times$ length] and $a$ is the unit distance between neighboring particles in our simulations.
By combining the expressions for the overall elastic modulus of the ladder model ($G_{0}$) as a function of the number $n$ of elements in each ladder structure (Eqs.~\eqref{laddertofrac} and \eqref{laddertofrac2}), we find that $G_{0}$ is related to the torsional stiffness $K_0$ through a structure-dependent factor $G_0\propto(L/\xi)^{d-2}K_0/(a^3n^{2\alpha})$. Similarly, in disordered elastic networks with bending elasticity \cite{Kantor1984}, the scaling $G_{0}\propto(L/\xi)^{d-2}K_\xi/L$ %with $K_{\xi}$ in elastic networks with bending interactions 
implies that $G_0\propto(L/\xi)^{d-2}K_0N_{\text{SCB}}^{-2\nu-1}a^{-2}/L$. These two distinct scaling expressions for $G_0$ suggest the following inter-relationship for the number of mechanical elements in each ladder structure
\begin{equation}\label{laddertofrac4}
   n\propto \left (\frac{L}{a}\right)^{1/2\alpha}N_{\text{SCB}}^{(2\nu+1)/2\alpha}.
\end{equation} 
Since each element of linear size $\propto \xi$ implies a number $N_{\text{SCB}}$ of singly connected bonds \cite{Kantor1984}, with each pairwise combination of these being a source of bending interactions, we hypothesize that $n\propto N_{\text{SCB}}^2$. Combined with Eq.~\eqref{laddertofrac4}, this %sets a constraint for 
constrains the power-law exponent in the ladder model
\begin{equation}\label{alpha-nu}
\alpha=(2\nu+1)/4
\end{equation}
Using $\nu \simeq 0.8$ yields $\alpha=0.68\pm0.1$ for our percolated gels in 3D. The data for $\tan \delta$ and the corresponding fits from both fractional and ladder models confirm this prediction 
[Figs.~\ref{fig2}(b) and \ref{fig2}(c)].

The product $\eta = G_0 \tau$ sets the large scale rate of dissipation in the gel and is found to be independent of the volume fraction of branching points [Fig.~\ref{fig4}(a)]. The viscous dashpot $\eta_0$ in the ladder model (which is linked to $\tilde{\eta_0}$, $n$ and $\eta$ by Eqs.(\ref{laddertofrac}) and (\ref{laddertofrac2})) therefore follows the scaling:  
\begin{equation}\label{laddertofrac3}
   \eta_0\propto\left(\frac{L}{a}\right)^{d-3+1/\alpha}N_{\text{SCB}}^{\nu d-3}\eta
\end{equation} 
where $\eta$ can be also directly connected to the drag coefficient $\zeta$ in our simulations (see Methods section). Thus, Eq.~\eqref{laddertofrac3} can be understood as a volumetric average measure of the viscous dissipation in a $d-$dimensional box of size $\xi$ that is filled with $N_{\text{SCB}}$ singly-connected bonds. 

Finally, using Eq.~\eqref{laddertofrac}, we can now directly connect the microscopic physics of the gels to the hierarchical organization of the mechanical elements in the ladder model. Indeed, asymptotic expansion of the recursive relations that specify the ladder model [Eq.~\eqref{recur} in Methods] produce power-law decays for both the elastic and viscous coefficients as a function of the mode number, i.e., we expect
$E_i\propto E_0/i^{2\alpha-1}$ and $\eta_i\propto \eta_0/i^{2\alpha-1}$. We show in the following that this hierarchy of internal modes has its origins in the geometrical distribution of effective bending coefficients within the fractal blobs. We first consider the effective bending stiffness that arises from the torsion around the equilibrium angle $\theta$ for a certain bond when a force $\delta F$ is applied on the $I$-th neighboring bond away from it along the elastically active backbone of the gel network [see sketch in Fig.~\ref{fig4}(b)]. Such an effective bending stiffness decreases by increasing the distance between the bonds along the backbone. As the relative neighboring distance varies $1\leq I\leq N_\text{SCB}$, the number of modes in our ladder model varies in the corresponding range $1\leq i\leq N_\text{SCB}^2$, suggesting that $I=i^{1/2}$ is a reasonable mapping between the $i^{th}$ relaxation mode in the ladder model and the portion of a blob constructed from $I$ bonds.
We show in the Methods section that the effective stiffness for mode $i=I^2$ in the ladder model scales as $k_{i=I^2}=\delta F/\delta x \propto K_0/(I^{2\nu}a^2)$, and that the equivalent spring modulus of the $i$-th mode 
is $E_{i=I^2}\propto (K_0/a^3)/I^{2\nu-1}=(K_0/a^3)/i^{\nu-1/2}$. An identical power-law decay in fact appears in the corresponding scaling for the viscous model parameters and one can clearly identify $(E_i,\eta_i)=(1/i^{\nu-1/2})(E_0,\eta_0)$. Using Eq.~\eqref{alpha-nu}, these two power-law decays simplify to $(E_i,\eta_i)=(1/i^{2\alpha-1})(E_0,\eta_0)$, which, remarkably, corresponds to the recursive relationship required in the ladder model to produce a power-law rheological response. These results demonstrate that the geometrical, self-similar arrangement of singly connected bonds and the cooperative dynamics of bending interactions within individual fractal blobs lie jointly at the origin of the hierarchical order of the corresponding ladder-based/fractional models that compactly and effectively capture the power-law response of different colloidal gels over a wide range of timescales. On this basis, the viscoelastic master curve becomes a {\it discriminating probe} of the proximity to the rigidity threshold and of the marginal stability of particulate gels.

The fractal units and hierarchy of connectivities embedded in soft particulate gels may be embedded in the static microstructure (clusters, strands, meshes, etc.) that is directly accessible through confocal imaging or scattering, but they are revealed by measurements of linear viscoelasticity because these hidden structures govern stress transmission and elasticity. As such, they are naturally akin to force chains in granular media, or localized excitations arising in amorphous solids \cite{Cates:1998,Nampoothiri2020emergent}, and their spatial organization potentially determines the hierarchical stress transmission and redistribution under load, from particles to clusters and strands \cite{Hsiao2012,lindstrom2012,vanDoorn2018strand}. The ideas presented here show how mechanical spectroscopy can be used to understand the emergent viscoelastic properties of a broad range of technologically relevant materials, providing insight across a broad experimental literature and a new scientific basis for material design in areas from 3D printing to recycling. Future work, in fact, can build on this study to investigate the implications of fractal characteristics and hierarchical organization of  particulate gels for non-linear properties, memory encoding, and smart adaptive response of soft materials.

\section*{Methods}
\textbf{Numerical model and simulation.} 
In the simulations, colloidal particles or aggregates, described as spherical objects of diameter $a$, spontaneously self-assemble into a gel network due to attractive short-range interactions ~\cite{israelachvili2015intermolecular}, of maximum strength $\varepsilon$, mediated by the solvent in which they are immersed and through which their thermal motion is overdamped. In real particulate gels, surface roughness, shape irregularity and sintering processes limit the relative motion of particles as they aggregate \cite{Dinsmore_2002,Dibble2008,Whitaker2019}. These effects are included in the model through an angular modulation of the net attraction that introduces a bending rigidity of the interparticle bonds \cite{Bantawa2021}. Each gel is characterized by its solid volume fraction estimated as $\phi = (\pi/6)Na^{3}/L^{3}$ with $N$ the total number of particles and $L$ the linear size of the cubic simulation box. For each value of $\phi$, various gel microstructures are obtained by tuning the rate $\Gamma$ at which the relative strength of the attractive interactions (with respect to $k_{B}T$) is increased to induce gelation during the sample preparation. For the set of model parameters used here, all networks start from one-particle thick semi-flexible strands (where particles have coordination number $z$=$2$) 
that branch ($z$=$3$) to reduce steric hindrances and frustration as they grow from different directions. Starting from these relatively simplified structural units, however, large scale numerical simulations ($> 10^{5}$ colloidal units) allow for hierarchical loops and larger scale heterogeneities to naturally emerge during the gel self-assembly, depending on the gelation rate. As a consequence, the resulting disordered and heterogeneous network topologies are  representative, at a coarse-grained level, of the structural complexity typical of a wide range of soft particulate gels \cite{Dibble2008,Whitaker2019,Laurati2014,Tsurusawa2019}.

For all $\phi$ and $\Gamma$ considered here, any further aging of the gels beyond the gel preparation is much slower than the simulation time window used to compute the rheological response of the samples; hence it can be considered negligible in the context of this study.
We use samples with $N$ varying between $2\cdot 10^3$ and $2\cdot 10^5$, $L$ varying from $23$ to $120$ particle diameters. $\Gamma$ is varied between $10^{-2}\epsilon/k_B\tau_0$ and $10^{-6}\epsilon/k_B\tau_0$ and the range of particle volume fractions spans $0.05 \leq \phi \leq 0.15$. In the model, particles interact via a short-range attraction $U_2$ and a three body term $U_3$ which introduces a bending stiffness between neighboring bonds. the Molecular Dynamics (MD) simulation with periodic boundaries is implemented for a system of $N$ particles in a cubic box of size $L$ with position vectors \{$\mathbf{r}_1,..., \mathbf{r}_N$\} and interacting with a potential energy:
\begin{equation}
U(\textbf{r}_1,..., \textbf{r}_N)=\varepsilon \bigg{[}\sum_{i>j}U_2\left(\frac{\textbf{r}_{ij}}{a}\right)+\sum_i\sum_{j>k}^{j,k\neq i}U_3\left(\frac{\textbf{r}_{ij}}{a},\frac{\textbf{r}_{ik}}{a}\right)\bigg]
\label{Pot}
\end{equation}
where $\textbf{r}_{ij}=\textbf{r}_j-\textbf{r}_i$. The functional forms of $U_2$ and $U_3$ are given in the SI (Section~11) and their detailed description can also be found in previous works \cite{Colombo:2014JOR,Bouzid:2018LAN,Bantawa2021}. The MD time unit is expressed in terms of particle mass $m$, diameter $a$ and unit energy $\varepsilon$ as $\tau_0=\sqrt{ma^2/\varepsilon}$. All other physical quantities are measured in units of $m$, $a$, $\varepsilon$ and $\tau_0$. The equation of motion is solved using a Verlet algorithm with a time step $\delta t=0.005 \tau_0$. All simulations have been performed using the open source software LAMMPS \cite{PLIMPTON:1995JCP} modified to incorporate the potential energy [Eq.~\eqref{Pot}].\\

\textbf{Gel Preparation.} 
The initial gel configurations are prepared by following the protocol described in \cite{Colombo:2014JOR,Bantawa2021}. Below, we briefly summarize the procedure:  We use NVT equilibrium MD simulations, with a Nosé-Hoover (NH) thermostat to cool down a system of particles in a gas phase initially at a reduced temperature $k_BT_i/\varepsilon=0.5$ down to $k_{B}T_f/\varepsilon =0.05$ in $N_\mathrm{cool}$ MD steps which define the cooling (or gelation) rate as $\Gamma = \Delta T / \Delta t = (T_f-T_i)/N_\mathrm{cool}\delta t$. We verify that the final temperature $k_{B}T_f/\varepsilon =0.05$ is low enough for the particles to aggregate and form a percolated gel network. Then, we let the system further equilibrate at $k_{B}T_f/\varepsilon = 0.05$ with the NH thermostat for another $N_\mathrm{equi}$ MD steps. To vary the gelation rate $\Gamma$, we change the number of MD steps used for cooling, i.e., $N_\mathrm{cool}$. The gel configurations are then obtained by draining the kinetic energy from the system which is carried out by quenching them to $k_BT/\epsilon \approx 0$ by using dissipative microscopic dynamics, which guarantees that the configurations are stuck in the local minimum of the potential energy, or inherent structure: 
\begin{equation}
m\frac{d^2\textbf{r}_i}{dt^2}=-\nabla_{\textbf{r}_i}U-\zeta\frac{d\textbf{r}_i}{dt}
\end{equation}
where $\zeta$ is the drag coefficient of the given solvent and we choose $m/\zeta=1.0\tau_0$.

The data for varying volume fraction $\phi$ correspond to a fixed gelation rate of $\Gamma=10^{-5} \varepsilon/k_B \tau_0$ with $N_\mathrm{cool}=10^6$ MD steps and $N_\mathrm{equi}=10^6$ MD steps. The data for varying gelation rates $\Gamma$ correspond to a fixed volume fraction $\phi=10\%$ and gelation rates vary in the range $\Gamma=10^{-6}-10^{-2} \varepsilon/k_B \tau_0$ with $N_\mathrm{cool}=10^8-10^4$ MD steps and $N_\mathrm{equi}=2\cdot10^4$ MD steps. This variation of gelation rates corresponds to a change in low frequency elasti moduli by an order of magnitude [cf. Fig 1(d)-(f)].\\

\textbf{Linear viscoelastic spectra.} For each gel, we use a computational scheme \cite{Bouzid2018} that has been inspired by a recently developed experimental technique \cite{Geri2018} and obtain the full linear viscoelastic spectrum by applying an optimally windowed chirp (OWCh) signal. The details of this protocol are presented in the SI (see Section~11).\\

\textbf{Master curve for the loss tangent}\\
As demonstrated in the Figure 2(a), the viscoelastic spectra of the low volume fraction gels, that are close to the percolation limit, follow the principle of time-connectivity superposition and the corresponding values of loss tangent will collapse on a master-curve for these gels. This simple collapse, which is obtained with just a horizontal shift of the data, is a signature of the self-similarity that exists between the shape of the measured viscoelastic spectra. Thus, we can think of our method (in generating a $\tan\delta$ master curve in Figure 2(a) from the measured spectra in Figure 1(g)) as a general ``discriminating probe" that determines whether the backbone of a mature gel is still similar to the original structure that was formed at the percolation or not.\\

\textbf{The mathematical arrangement of model parameters in ladder models.}
As discussed in \cite{Schiessel1993,Schiessel1995c}, with simple analysis of continued fractions for ladder models one can show that the following recursive arrangement is required to obtain a power-law behavior for the viscoelastic moduli that approaches the critical gel behavior with exponent $\alpha$:  
 \begin{equation}
\begin{split} \label{recur}
    & \tilde{E_i}=\frac{1}{2i-1}\frac{\Gamma(\alpha)}{\Gamma(1-\alpha)}\frac{\Gamma(i+1-\alpha)}{\Gamma(i-1+\alpha)}\tilde{E_0}\\
    & \tilde{\eta_i}=2\frac{\Gamma(\alpha)}{\Gamma(1-\alpha)}\frac{\Gamma(i+1-\alpha)}{\Gamma(i+\alpha)}\tilde{\eta_0}.
\end{split} 
 \end{equation} 
where $\Gamma$ is the complete Gamma function and $1~\leq~i~\leq~n$ represents the parameter index in the ladder model. This arrangement of parameters leads to a frequency-independent regime for the loss tangent that spans the frequency range  $(1/n^2)E_0/\eta_0\leq \omega \leq E_0/\eta_0$ . For frequencies smaller and larger than the specified span, the ladder model displays asymptotic single-mode retardation and single-mode relaxation consistent with the predictions of the Kelvin-Voigt and Maxwell models respectively.\\
By fitting the viscoelastic spectra with the proposed ladder model, one can find an estimate for the number of ladder elements $n$ and develop a quantitative measure for the range of scale-free relaxation modes in a given gel system. This can be connected to the difference between bounding cut-off length-scales and time-scales in the underlying fractal structure and relaxation spectra of real networks (see section 2 and Figure S2 of the SI for further details).  Similarly, as shown in Equation 10, by determining power-law exponent $\alpha$ of the ladder or fractional model we gain extra insight into the hierarchical arrangement of springs and dashpots in the mechanical ladder structure.\\

\textbf{Microscopic dynamics.}
The microscopic dynamics of the gel are probed by using a Langevin dynamics:
\begin{equation}\label{LangevinEqn}
m\frac{d^2\textbf{r}_i}{dt^2}=-\nabla_{\textbf{r}_i}U-\zeta\frac{d\textbf{r}_i}{dt}+F_r^i (t)
\end{equation}\\
where $m/\zeta=10.0\tau_0$ and $F_r^i (t)$ is a random white noise that mimics thermal fluctuations and is related to the drag coefficient $\zeta$: $\langle F_r^i(t) F_r^j(t^\prime)\rangle=2\zeta k_B T\delta_{ij} \delta(t-t^\prime)$. From the spontaneous particle dynamics with thermal fluctuations $k_BT/\varepsilon=10^{-3}$ (large enough to induce particle motion but without changing the topology), we monitor the position of particle $i$ at time $t$; ${\bf{r}}_i(t)\equiv(x(t),y(t),z(t))$, where $x$, $y$ and $z$ represent the Cartesian coordinates. The magnitude of the gel displacement is computed as  $\Delta_i=||{\bf{r}}_i(t_0+t_w)-{\bf{r}}_i(t_0)||$. Both the initial time $t_0$ and waiting time $t_w$ are chosen to be in the plateau in the particle mean-squared displacement curve [see Fig.~S4] with $t_0=10^4 \tau_0$ and additional waiting time $t_w=10^4 \tau_0$.\\

\textbf{Scaling of Physical Distance and Deformations in the Fractal Element.}
Due to the fractal nature of the blobs, the physical distance between neighbors that are $I$-bonds apart is $d\propto a I^\nu$ and the corresponding torque from force $\delta F$ scales as $\delta FI^\nu a$, which leads to a change of angle $\delta \theta \propto \delta FI^\nu a/K_0$. The rotation of this angle leads to a local deformation of $\delta x\propto aI^{\nu}\delta \theta$ in direction of the applied force $\delta F$. \\

\textbf{Persistence length.} The persistence length $l_p$ is determined by computing the correlation in the angles of successive bonds along the strand \cite{Bantawa2021} of a $50$-particle strand sampled over different configurations from the dynamics [Eq.~\eqref{LangevinEqn}] at a finite temperature. For the potential parameters used in this study, the persistence length is estimated to be $l_p$  $\sim5.5a$.

\section*{References}
\bibliography{main}

\begin{acknowledgments}
M.Bantawa and E.D.G. acknowledge financial support from Georgetown University and National Science Foundation (NSF DMR-2026842). This research was supported in part by the National Science Foundation under Grant No.~NSF PHY-1748958 through the KITP program on the Physics of Dense Suspensions.
\end{acknowledgments}
\section*{AUTHOR CONTRIBUTIONS}
M. Bantawa and B.K. contributed equally to this work; M. Bantawa performed simulations; M. Bantawa, B.K., M.G., M. Bouzid, T.D., G.H.M. and E.D.G. analyzed data; and M. Bantawa, B.K., M.G., M. Bouzid, T.D., G.H.M. and E.D.G. wrote the paper.

\clearpage
\setcounter{figure}{0} 
\renewcommand{\thefigure}{S\arabic{figure}}
\renewcommand\theequation{S\arabic{equation}}    
\setcounter{equation}{0}
\renewcommand\thesubsection{S.\arabic{subsection}}

\section*{Supplementary information}
\begin{figure*}[htb]
\centering
\includegraphics[scale=0.8]{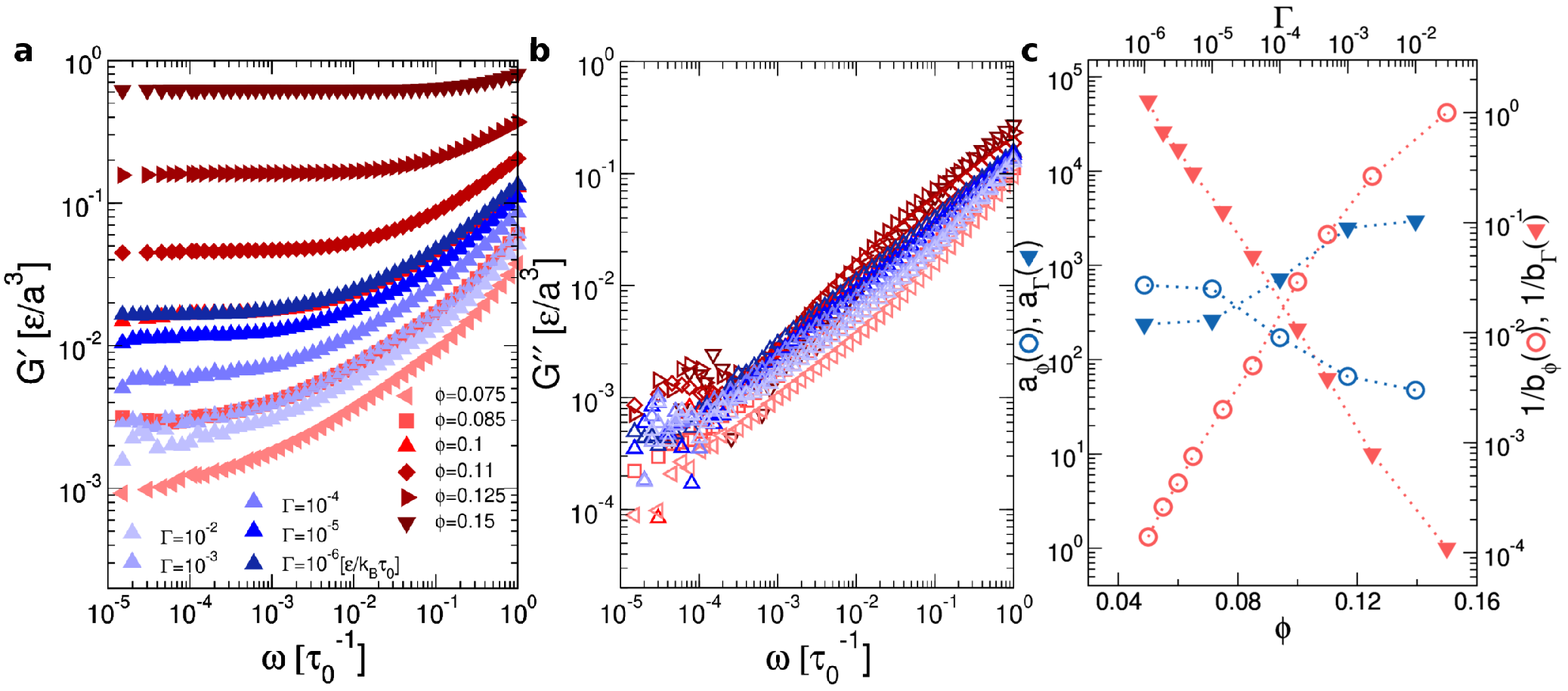}
\caption{\textbf{Linear viscoelastic spectra of gel networks obtained by varying the volume fraction $\phi$ and the gelation rate $\Gamma$.} The storage modulus $G^\prime$ in (a) and loss modulus $G^{\prime \prime}$ in (b) are shown as a function of angular frequency $\omega$ for gels with different volume fractions $\phi$ and gelation rates $\Gamma$. For gels with different volume fractions, the gelation rate is fixed at $\Gamma=10^{-5}\epsilon/k_B\tau_0$.  The gels with different gelation rates correspond to a fixed volume fraction $\phi=0.1$. (c) The horizontal shift factors (left axis), $a_\phi$ (red triangles) and $a_\Gamma$ (blue triangles), and the vertical shift factors (right axis), $b_\phi$ (red circles) and $b_\Gamma$ (blue circles), are plotted as functions of the volume fraction $\phi$ (bottom axis) and gelation rate $\Gamma$ (top axis). The dotted lines are guides to the eye.
 }\label{fig1SI}
\end{figure*}
\subsection{Linear viscoelastic spectra}
The storage and loss moduli of gels obtained by varying the volume fractions $\phi$ and gelation rate $\Gamma$ are shown in Fig.~\ref{fig1SI}. These data were rescaled to construct the master curves in Fig.~2(b) shown in the main text. The storage moduli vary by approximately three orders of magnitude.
\subsection{Fractional model}
 Fractional models are often used as concise mechanical representations of soft gelling systems\cite{WinterChambon1986,Jaishankar:2013, KeshavarzPNAS2021}. The simplest fractional model, known as the spring-pot, was introduced by Scott-Blair \cite{Blair1939} to capture the scale-free power-law behavior that is often observed in soft gels and foodstuffs. A single spring-pot predicts power-law behavior for both loss and storage moduli and, consequently, a frequency-independent loss tangent. The fractional model that we used in our study is a modified Fractional Kelvin-Voigt (FKV) model \cite{Jaishankar:2013} that consists of a spring element $G_0$ that is connected in parallel to a dashpot $\eta$ and springpot ($\mathbb{V},\alpha$) combination. The complex moduli $G^*=G'+iG''$ for this model is:
 \begin{equation}\label{FKV}
    G^*(\omega)=G_0\left[1+(i\omega\tau)\frac{1}{1+(i\omega\tau_l)^{1-\alpha}}\right]
 \end{equation}
where $\tau=\eta/G$ is the retardation timescale and $\tau_l=(\eta/\mathbb{V})^{1/(1-\alpha)}$ is the cut-off timescale that captures the transition from faster power-law relaxation modes to a slower single-relaxation terminal mode. At low frequencies, which correspond to long timescales, the model predicts a behavior similar to a single-mode Kelvin-Voigt solid  where $G'(\omega)\sim G$ and $G''(\omega)\propto \eta\omega$. At high frequencies, which correspond to short timescales, the model predicts power-law behavior for both loss and storage moduli $G'\sim G''\propto \omega^\alpha$, which is identical to the behavior observed in critical gels in the vicinity of the gelation point. This scale-free behavior for shorter modes captures the hidden fractal nature of mechanical relaxation that exists within individual clusters.

The fact that we collapsed all the measured spectra for the loss tangent onto a single master curve with only one horizontal shift factor suggests that there is one important characteristic time scale $\tau_l\sim\tau$. In other words, we conclude that in our studied system the value of the quasi-property is set by other model parameters $\mathbb{V}\propto G (\eta/G)^\alpha$, which is self-consistent with the physical analogy to a ladder model in which the quasi-property or scale factor controlling the magnitude of the stress naturally emerges as a combination of the individual spring moduli $E_i$ and dashpot viscosities $\eta_i$. An identical relationship was discussed by Bagley and Torvik \cite{Bagley1983} in their derivation of fractional rheology from multi-mode Rouse relaxation processes in polymer solutions. We have also obtained a similar constraint in our previous numerical study of colloidal gels \cite{Bouzid2018}.
\begin{figure}[htb]
\centering
\includegraphics[width=0.52\textwidth]{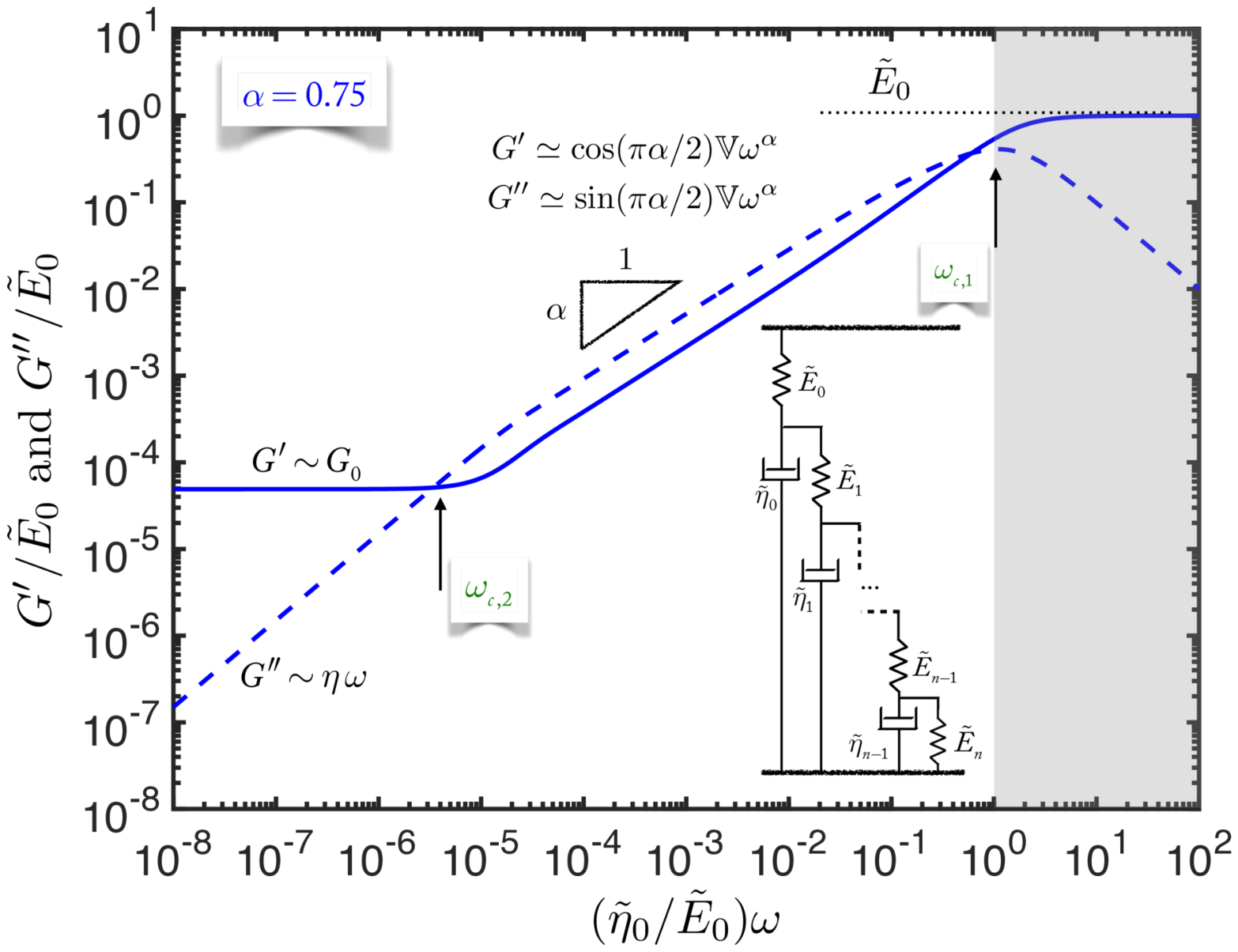}
\caption{\textbf{Rheological predictions of the ladder model.} The viscoelastic storage and loss moduli are calculated for a ladder model that consists of $n=5000$ elements, where the model parameters follow the prescribed arrangement given in Equation \ref{recur}. Solid and dashed lines represent elastic and loss moduli respectively. The shaded area shows the high-frequency region $(\omega_{c,1}=\tilde{E}_0/\tilde{\eta}_0)\leq\omega$ where we observe a single-mode relaxation at timescales shorter than the smaller cut-off timescale of the model. This high-frequency behavior of the model is similar to the simple Maxwell model and is beyond the probed range of frequencies in our simulations. At low frequencies $\omega\leq (\omega_{c,2}=(1/n^2)\tilde{E}_0/\tilde{\eta}_0)$, the model predicts a single-mode retardation behavior at large timescales that resembles the behavior of a simple Kelvin-Voigt element. For intermediate frequencies $\omega_{c,2}\leq \omega\leq \omega_{c,1}$ the ladder model displays a scale-free power-law behavior for both elastic and loss moduli that is similar to the rheological response of a polymer gel at the gelation point.}
\label{LadderModelRheology}
\end{figure}

\subsection{Ladder Models and Connection to the Fractional Model}
As discussed in Refs.~\cite{Schiessel1993,Schiessel1995} and also mentioned in the main text, ladder models can capture the power-law behavior observed in our computation of the viscoelastic moduli. To obtain a power-law behavior for viscoelastic moduli that resembles colloidal gel behavior with exponent $\alpha$ close to the gel threshold (critical gel), we require the following arrangement for the model parameters:  
 \begin{equation}
\begin{split}
    & \tilde{E_i}=\frac{1}{2i-1}\frac{\Gamma(\alpha)}{\Gamma(1-\alpha)}\frac{\Gamma(i+1-\alpha)}{\Gamma(i-1+\alpha)}\tilde{E_0}\\
    & \tilde{\eta_i}=2\frac{\Gamma(\alpha)}{\Gamma(1-\alpha)}\frac{\Gamma(i+1-\alpha)}{\Gamma(i+\alpha)}\tilde{\eta_0}.\label{recur}
\end{split} 
 \end{equation} 
where $\Gamma$ is the Gamma function and $1\leq i \leq n$ represents the parameter index in the ladder model. As shown in Fig.~\ref{LadderModelRheology}, this arrangement of parameters leads to a frequency-independent regime for the loss tangent that spans the range  $(1/n^2)E_0/\eta_0\leq \omega \leq E_0/\eta_0$ range. For frequencies smaller and larger than the specified span, the model displays a single-mode retardation and a single-mode relaxation time that resemble the predictions of Kelvin-Voigt and Maxwell models respectively.

The rheological response of the ladder model can be divided into three different regions:
\begin{itemize}
    \item Low-frequency response: for frequencies below a critical value $\omega_{c,2}=(1/n^2)\tilde{E}_0/\tilde{\eta}_0$, we observe a behavior similar to a single Kelvin-Voigt model where $G'\sim G_0={\tilde{E}_0}/{n^{2\alpha}}$ and $G''\propto\eta \omega=n^{2-2\alpha}\tilde{\eta}_0 \omega$.
    \item Medium-frequency response: for intermediate frequencies $\omega_{c,2}\leq \omega\leq \omega_{c,1}=\tilde{E}_0/\tilde{\eta}_0$, we observe a scale-free power-law behavior that is similar to the predictions of a simple springpot for a critical gel with exponent $\alpha$ and quasi-property $\mathbb{V}=\tilde{E}_0(\tilde{\eta}_0/\tilde{E}_0)^\alpha$. 
    \item High-frequency response: for large frequencies $(\omega_{c,1}=\tilde{E}_0/\tilde{\eta}_0)\leq\omega$, we observe a single-mode relaxation at timescales shorter than the smallest cut-off timescale of the model. This high-frequency behavior of the model is similar to the simple Maxwell model and is beyond the probed range of frequencies in our simulations. 
\end{itemize}
By direct comparison between the modified FKV model (Eq.~\ref{FKV}) and the low and intermediate frequency regions of the ladder model, we conclude that both models predict similar rheological behavior for the viscoelastic moduli and that the corresponding parameters of the modified FKV model are related to the ladder model by the following relationships:
\begin{equation}\label{laddertofrac}
   G_0=\frac{\tilde{E_0}}{{n}^{2\alpha}},\:\eta={n}^{2-2\alpha}\tilde{\eta_0},\:\mathbb{V}=\tilde{E_0}\left(\frac{\tilde{\eta_0}}{\tilde{E_0}}\right)^\alpha.
\end{equation}
Using Equations \ref{recur} and \ref{laddertofrac}, we can use the fitting parameters of the ladder model (or fractional model) in an "inverse" manner and learn more about important hidden features in the mechanical relaxation of the underlying structure. The number of ladder elements $n$ provides a quantitative measure for the range of scale-free relaxation modes in a given gel system, which can be connected to the difference between bounding cut-off length-scales and time-scales in the underlying fractal structure and relaxation spectra of real networks.  Similarly, by determining power-law exponent $\alpha$ of the ladder or fractional model we gain extra insight into the hierarchical arrangement of springs and dashpots in the mechanical ladder structure.
\subsection{Scaling discussion for the volume fraction of the branching points}
Many studies have reported varying scaling exponents for the evolution of the gel modulus with volume fraction. Our model reveals that these apparent differences in the literature may simply be due to the fact that the volume fraction $\phi$ is not the best measure for distance from the rigidity percolation threshold and indeed a better candidate can be found by studying the volume fraction of the branching points $\phi_{\text{br}}$. Figure 3(d) and Table 2 in the main manuscript help the reader to see how different aggregation processes show different exponents for the evolution of the gel modulus with volume fraction. It is however  noteworthy to mention that the measuring volume fraction of the branching points $\phi_{\text{br}}$ can be challenging in many experimental studies. As mentioned in the main manuscript the following scaling helps us to compute the volume fraction of the branching points $\phi_{\text{br}}$ from values of particle volume fraction and measured fractal dimensions:
\begin{equation}
 \phi_{\text{br}}\propto\phi^{1/\nu(d-d_f)}   
\end{equation}
where $d=3$ is the Euclidean dimension of a 3-D space and $d_f$ is the fractal dimension of the network inside the fractal clusters. According to \cite{Stauffer2018}, the computed estimate for $\nu \simeq 0.80 \pm 0.16$ is compatible with the value of a 3D random percolation network.\\ We have also shown that the evolution of the gel modulus with volume fraction follows a power-scaling:
\begin{equation}
 G_0\propto\phi^f_{\text{obs}} 
\end{equation}
where $f_{\text{obs}}=f/\nu(d-d_f)$ and $f\simeq 3.5$ according to Kantor and Webman \cite{Kantor1984}. We can use this scaling in an inverse manner. One can simply measure the elastic moduli at different volume fraction and find an accurate estimate for $f_{\text{obs}}$, from which we can find an estimate fro the fractal dimension of the underlying network inside the clusters: 
\begin{equation}
    d_f=d-\frac{1}{\nu}\frac{f}{f_{\text{obs}}}.
\end{equation}

\subsection{Box counting and fractal dimension}
\begin{figure}[htb]
\centering
\includegraphics[width=0.45\textwidth]{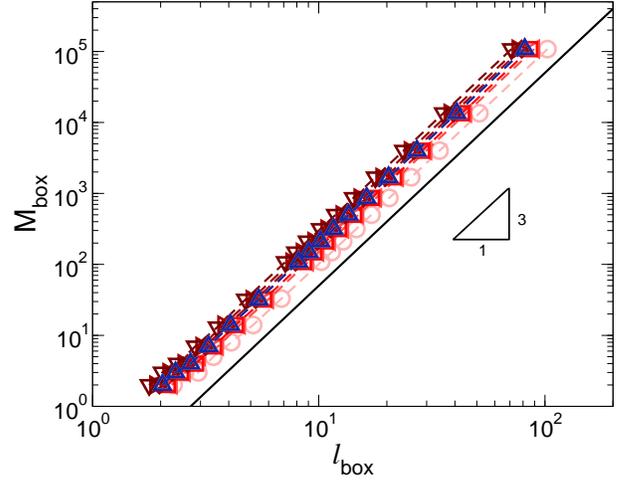}
\caption{\textbf{Demonstration of the non-fractal nature of gel networks.}
The mass in the sub-box $M_{\rm box}$ as a function of $l_{\rm box}$. The symbols and colors are the same as in Fig.~\ref{fig1SI}.}
\label{BoxCounting}
\end{figure}

We employed box-counting method to check the fractality of the gel structures. Each simulation box is divided into sub-boxes of size $l_{\rm box}$ and the mass in each box $M_{\rm box}$ is computed. The data showing $M_{\rm box}\propto (l_{\rm box})^3$ in Fig.~\ref{BoxCounting} indicates that the gel structures are non-fractal.

\subsection{Microscopic particle dynamics}
The mean-squared displacement (MSD) of particles in the gel networks is computed from the microscopic trajectories obtained by solving the Langevin equations of motion with thermal fluctuations $k_BT/\epsilon=10^{-3}$ and is given by:
\begin{equation}
\langle \Delta r^2 \rangle (t)=\bigg \langle \frac{1}{N }\sum_{i=1}^N \big(\bf{r}_i(t)-{\bf{r}}_i(0)\big)^2 \bigg \rangle
\end{equation}
where ${\bf{r}}(t)\equiv(x(t),y(t),z(t))$ represents the particle coordinates at a given time $t$.

\begin{figure*}[htb]
\centering
\includegraphics[scale=0.95]{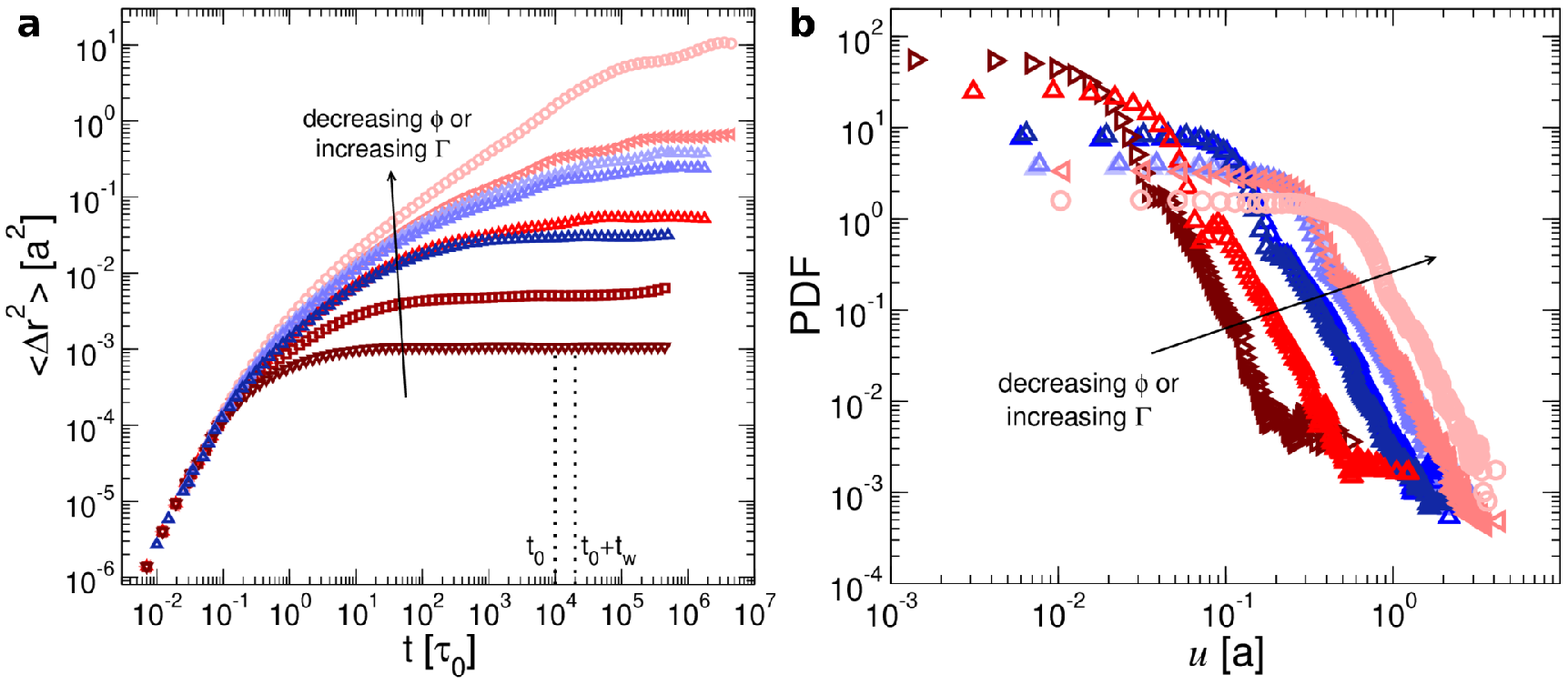}
\caption{\textbf{Microscopic particle dynamics.} (a) The mean squared displacements as a function of time.  (b) Probability distributions of fluctuations of displacements. The symbols and colors are consistent with Fig.~\ref{fig1SI}.}\label{figDynamics}
\end{figure*}
The particle displacements are computed with respect to a reference configuration in the MSD plateau (see Fig.~\ref{figDynamics}(a)), i.e., $t_0\simeq 10^4\tau_0$ as ${\bf \Delta}={\bf{r}}(t_0+t_w)-{\bf{r}}(t_0)$. 
From the PDF of the displacements, we then compute the fluctuations $u=[({\bf{\Delta}} - \langle {\bf \Delta} \rangle)^2]^{1/2}$ to characterize the particle dynamics that results from the heterogeneity of the gel structures \cite{Bantawa2021}. 
\begin{figure*}[htb]
    \centering
    \includegraphics[width=0.95\textwidth]{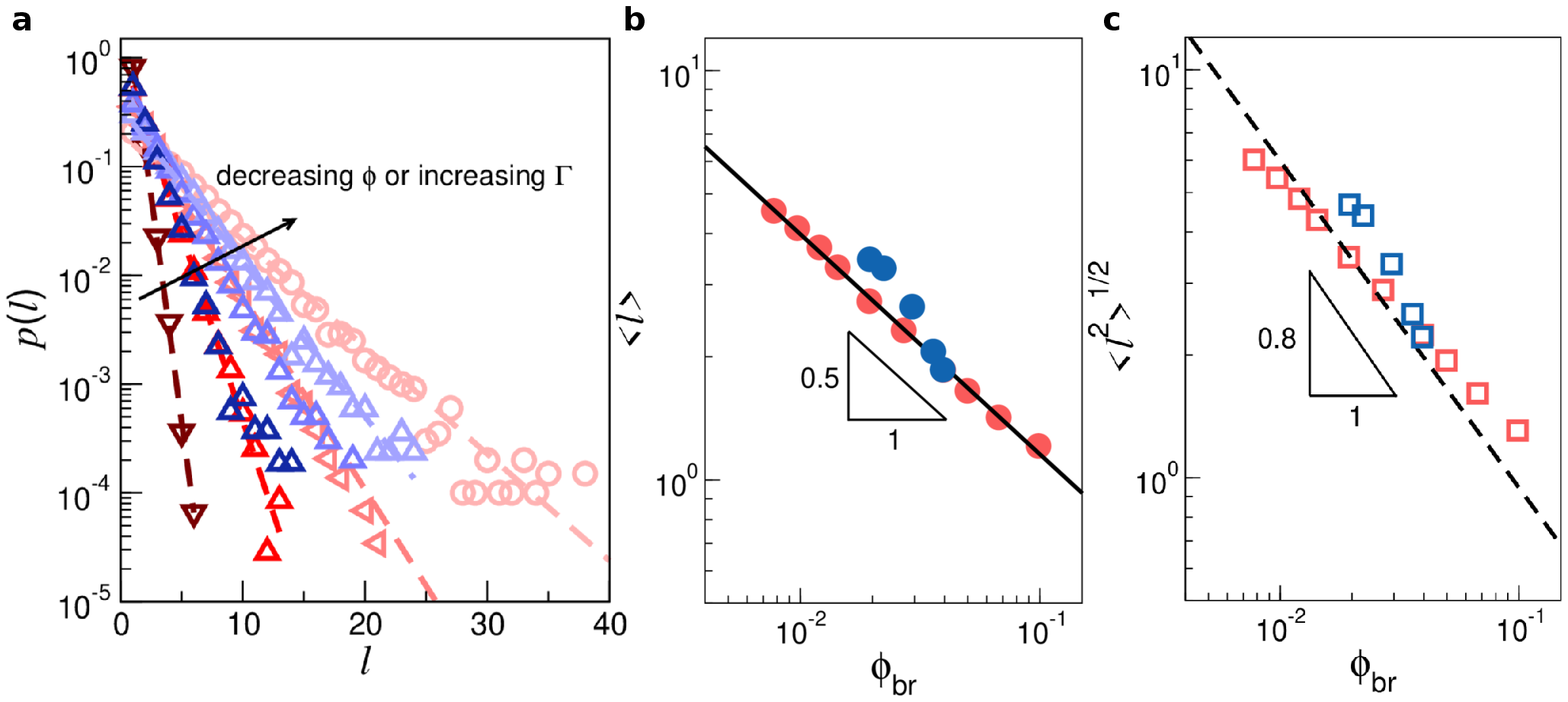}
    \caption{\textbf{Mesh size $l$ and relevant length scales in the gel networks.} (a) The probability distribution $p(l)$ of mesh sizes $l$, where $l$ is computed as the topological distance between two connected branching points for different levels of connectivity.
    Dashed lines correspond to the best exponential fits of the data. The symbols and colors are the same as in Fig.~\ref{fig1SI}. In (b) and (c) the length scales computed from the first and second moment of the mesh size distribution, i.e., $\langle l \rangle$ and $\langle l^2\rangle^{1/2}$, as a function of the volume fraction of branching points ($\phi_{\rm br}$). The red data correspond to different volume fractions $\phi$ while the blue data correspond to different gelation rates $\Gamma$. The solid line in (b) and dashed line in (c) represent power laws of exponent $-0.5$ and $-0.8$ respectively.}
    \label{Mesh size}
\end{figure*}

\begin{figure*}[htb]
\centering
\includegraphics[width=0.9\textwidth]{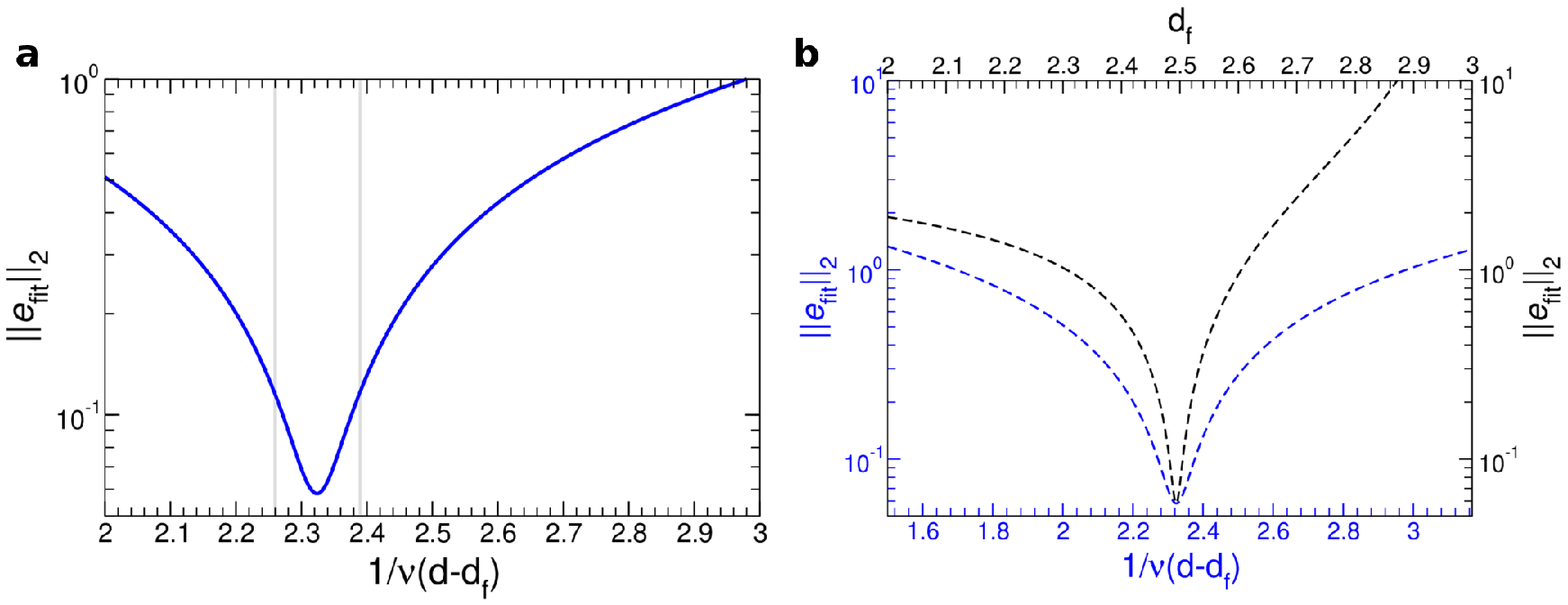}
\caption{\textbf{Estimation of error bar in the scaling of $\phi_{\rm br}$ vs. $\phi$.} (a) and (b) The $L_2$ norm magnitude of the fitting error for the power-law fit that we used in Figure 3(d) in the main text. As clearly demonstrated, the concept of Normal Equation in linear regression helps us to find the optimum values of fitting parameters that minimize the fitting error for a corresponding tall system of linear equations \cite{Strang2016}. We can thus select a confidence zone and report the optimum value with corresponding values of error bars $1/\nu(d-d_f)\simeq 2.31\pm 0.06$ [gray lines in (a)].}
\label{figS4}
\end{figure*}
\subsection{Mesh-size distribution}
As mesh size, we compute the topological distance between two neighboring branching points along the network. For all gels, the distributions of mesh sizes become wider with decreasing $\phi$ and increasing $\Gamma$. 
While these distributions can be at first sight fitted with an exponential form [Fig.~\ref{Mesh size}(a)], a more detailed analysis reveals that their first and second moment follow a different scaling with $\phi_{br}$ [Fig.~\ref{Mesh size}(b) and (c)], indicating a deviation from the exponential form at large $l$, which is however easily masked by the finite size of the simulations. The second moment $\langle l^2\rangle^{1/2}$ has a stronger dependence on $\phi_{br}$, indicating that the deviation from the exponential distribution must dominate close to the rigidity transition. These findings suggest that the critical correlation lengthscale associated with the rigidity transition may be related to the second moment $\langle l^2\rangle^{1/2}$, and to extract the relevant scaling we therefore use the root mean square fluctuations $\xi=\langle ( l - \langle l \rangle)^2\rangle^{1/2}$ [Fig.~3(c)], which are also less affected by finite size effects. The data for the lengthscale obtained from $\langle l^2\rangle^{1/2}$ [Fig.~\ref{Mesh size}(c)] indeed approach the same scaling as $\xi$ [Fig.~3(c)].

\subsection{Estimation of error bars in the exponent for dependence of $\phi_{\rm br}$ on $\phi$}
The method to obtain the error bar in the scaling $\phi_{\rm br}\propto \phi^{1/\nu (d-d_f)}$ is shown graphically in Fig.~\ref{figS4}.
\subsection{Fractal nature of percolating backbone}
\begin{figure*}[htb]
\centering
\includegraphics[scale=0.8]{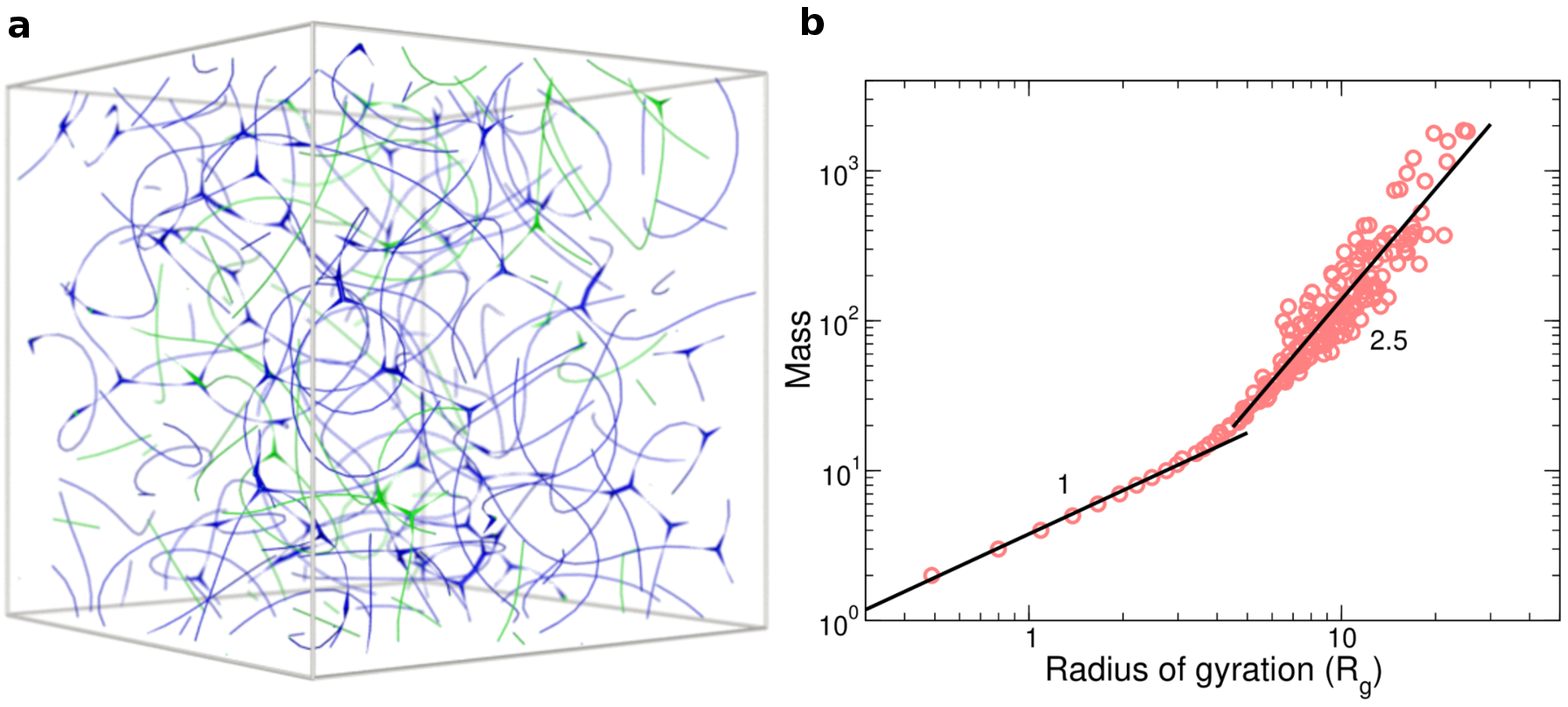}
\caption{\textbf{Characterization of fractal structure of the first percolation backbone near the gelation threshold.} (a) A simulation snapshot of the gel structure in the early stage of gelation. The blue network represents the percolating cluster (backbone) while the finite clusters, which are separated from the backbone, are shown in green. The thicker regions correspond to regions where stresses are accumulated as the gel solidifies. (b) Scaling of mass of the clusters as a function of their radius of gyration. From these data, we obtain the fractal dimension $d_f=2.5\pm 0.05$.}\label{figbackbone}
\end{figure*}

We have studied the fractal nature of the gel close to the gelation threshold by performing simulations at high temperature, where the gels are first formed [see the snapshot in Fig.~\ref{figbackbone}(a)]. The fractal dimension $d_f=2.5\pm 0.05$ [Fig.~\ref{figbackbone}(b)] can be obtained from the scaling of the mass of finite clusters as a function of their radius of gyration $R_g$.

\subsection{Measurements of viscoelastic parameters from microscopic dynamics}
\begin{figure}[htb]
\centering
\includegraphics[scale=0.8]{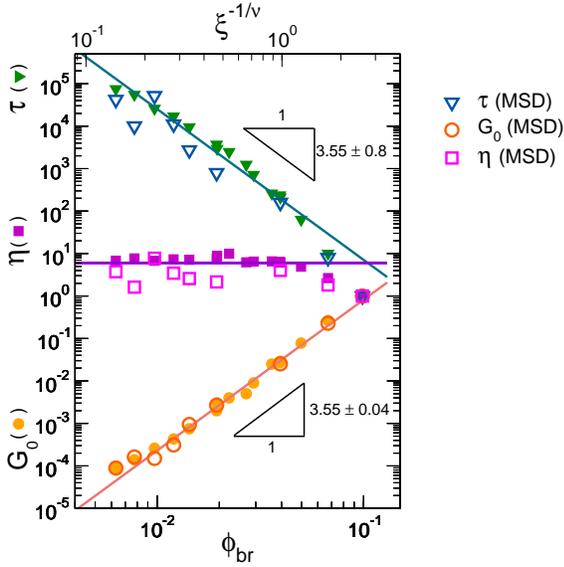}
\caption{\textbf{Comparison of viscoelastic parameters obtained from rheology and microscopic dynamics.} The filled symbols correspond to the viscoelastic parameters measured from the macroscopic rheology while the open symbols are the values obtained from the corresponding MSD.
 }\label{figParaDynamics}
\end{figure}

In Fig.~\ref{figParaDynamics}, we compare the retardation time $\tau$ obtained from the rheology [Fig.~4(a) in the main text] with the timescale $\tau(\textrm{MSD})$, where the $\textrm{MSD}$ reaches a plateau [see Fig.~\ref{figDynamics}(a)] and is obtained by locating the position of the minimum of the logarithmic-derivative of MSD. These two timescales have the same dependence on $\phi_\textrm{br}$ and $\xi$.  The elastic modulus $G_0$ obtained from the viscoelastic spectra and the estimated modulus $G_0(\textrm{MSD})$, obtained from the MSD value at the plateau also show an identical dependence on $\phi_\textrm{br}$ and $\xi$. 

\subsection{Gel Model}
The model used for the gel networks consists of monodisperse particles of diameter $a$, which represent colloidal particles or aggregates in real systems and spontaneously self-assemble into a gel network due to attractive short-range interactions\cite{israelachvili2015intermolecular}.
The short-range attractive potential $U_2$ is a combination of a repulsive core and a narrow attractive well, and is written in the following form, for computational convenience:
\begin{equation}
U_2(\textbf{r})=A \left(\frac{1}{r^{18}}-\frac{1}{r^{16}}\right)
\end{equation}
where $r$ is rescaled by the particle diameter $a$ and $A$ is the dimensionless parameters that controls the depth of the potential well. 

In particulate gels, the particle surface roughness or the irregular shapes of the aggregates can result in a significant hindrance of the relative particle motion when particles or aggregates are bonded in the gel state \cite{Dinsmore_2002,Dibble2008,Whitaker2019}. To include the energy costs associated with the constraints of the particle relative motion imposed by the nature of the surface contacts, we use a three-body term $U_3$. For two particles both bonded to a third one and whose relative positions with respect to it are represented by the vectors $\textbf{r}$ and $\textbf{r'}$ originating from the same particle, $U_3$ is expressed, again for computational convenience, in the following form:
\begin{equation}
U_3(\textbf{r},\textbf{r'})=B \Lambda({r})\Lambda({r'})\exp\left[-\left(\frac{\textbf{r}\cdot\textbf{r'}}{rr'}-\cos \theta\right)^2 w^{-2}\right]
\end{equation}
where $B$, $\theta$ and $u$ are dimensionless parameters. The range of three-body interaction is set to two particle diameters, as ensured by the radial modulation:
\begin{equation}
\Lambda(r)= r^{-10}\left[1-(r/2)^{10}\right]^2 {\mathcal{H}}(2-r)
\end{equation}
where ${\mathcal{H}}$ is the Heaviside function. The dimensionless parameters of the potential energy in $U_2$ and $U_3$ are adjusted to the following values: $A=23$, $B=67.27$, $\theta=65^{\circ}$ and $w=0.3$ such that a disordered and thin percolating network starts to self-assemble at $k_B T/\epsilon\approx 0.05$ \cite{Colombo:2014SM,Colombo:2014JOR,Bouzid2018,Bantawa2021}.

\subsection{Optimally Windowed Chirp (OWCh)}
The input chirp signal has a mathematical form:
\begin{equation}
\gamma (t)=\gamma_0 W(t;b) \sin [(L\omega_1) (e^{t/L}-1)]
\end{equation}
where $\gamma_0$ is the strain amplitude and $L = T / \ln(\omega_2/\omega_1)$. The phase of the signal exponentially grows from the initial frequency $\omega_1$ to the final frequency $\omega_2$ within the duration of the signal $T$. The window function $W(t;b)$ for a symmetric Tukey window is given by:
\begin{equation} \label{Window Function}
  W(t;b)=\left\{
  \begin{array}{@{}ll@{}}
    \cos^2[\frac{\pi}{b}(\frac{t}{T}-\frac{b}{2})], & \text{for}\ \frac{t}{T} \leq \frac{b}{2} \\
       1, & \text{for}\ \frac{b}{2} < \frac{t}{T} < 1-\frac{b}{2} \\
          \cos^2[\frac{\pi}{b}(\frac{t}{T}-1+\frac{b}{2})], & \text{for}\ \frac{t}{T} \geq 1-\frac{b}{2} \\
  \end{array}\right.
\end{equation}
The OWCh signal $\gamma(t)$ is used to impose deformation. Following each deformation, the equation
\begin{equation}\label{oscillatory}
 m\frac{d^2{\textbf{r}}_i}{dt^2}=-\nabla_{{\textbf{r}}_i}U - \zeta\big(\frac{d{\textbf{r}}_i}{dt}-\dot{\gamma}(t)y_i{\textbf{e}}_x\big)
\end{equation}
is solved for each particle using Lees-Edwards boundary conditions, where $\textbf{e}_x$ is a unit vector along the $X$- axis. We use $m/\zeta=10.0\tau_0$ to be in the overdamped regime. We compute shear stresses, $\sigma_{xy}(t)$ from the interaction part of the global stress tensor using the standard virial equation while neglecting the contribution from kinetic energy and viscous dissipation as:
\begin{equation}
\sigma_{\alpha \beta} = \frac{1}{L^3}\sum_{i}\frac{\partial U}{\partial r_i^{\alpha}}r_i^{\beta}
\label{vstress}
\end{equation}
where $\alpha$ and $\beta$ stand for Cartesian components \{$x,y,z$\}. The viscoelastic moduli are then extracted as:

\begin{equation}
\begin{split}
&G'(\omega_i)={\rm Re}\left\{\frac{\tilde{\sigma}(\omega_i)}{\tilde{\gamma}(\omega_i)}\right\}\\
&G''(\omega_i)={\rm Im}\left\{\frac{\tilde{\sigma}(\omega_i)}{\tilde{\gamma}(\omega_i)}\right\}
\label{FFT}
\end{split}
\end{equation}
where $\tilde{\sigma}$ and $\tilde{\gamma}$ are the Fourier transforms of the stress and strain signals respectively.

\end{document}